\documentclass[14pt, nofootinbib, preprint,
doublespace
]{revtex4}

\usepackage{graphics}
\usepackage[title]{appendix}
\usepackage[pdftex]{graphicx}
\usepackage{booktabs}
\usepackage{amsmath,amssymb}
\usepackage{subcaption}
\usepackage{graphicx}
\usepackage{caption}
\usepackage{mathtools}
\usepackage{tikz}
\usetikzlibrary{shapes.geometric}

\DeclareMathOperator{\EX}{\mathbb{E}}

\begin{document}

\title{
Gas Fees on the Ethereum Blockchain: From Foundations to Derivatives  Valuations
}

\author{Bernhard K  Meister \& Henry C W Price$^\dagger$}

\affiliation{ 
     $^\dagger$Centre for Complexity Science, 
      Imperial College London,   UK  
    }
     
\date{\today }

\begin{abstract}
\noindent
 The `gas fee', paid for inclusion in the blockchain,
 is analyzed in two parts. First, we consider how  
 `effort' in terms of resources required to process and store a transaction turns into a `gas limit', which, through a fee, comprised of the `base' and `priority fee' in the current version of Ethereum, is converted into the cost paid by the user. We adhere 
 closely to the Ethereum protocol to simplify the analysis and to constrain the design choices when considering  `multidimensional gas'.
 
Second,  we assume that the `gas' price is given `deus ex machina' by a fractional Ornstein-Uhlenbeck process and evaluate various derivatives.
These contracts can, for example, mitigate gas cost volatility. 
The ability to price and trade `forwards' besides the existing `spot' inclusion into the blockchain could 
enable users to hedge against future cost fluctuations.

Overall, this paper offers a comprehensive analysis of gas fee dynamics on the Ethereum blockchain, integrating supply-side constraints with demand-side modelling to enhance the predictability and stability of transaction costs.
 
\end{abstract}




\thanks{
bernhard.k.meister@gmail.com, hp1110@imperial.ac.uk
}

\maketitle












\noindent 


 \section{Introduction}
 \noindent
The supply and demand of blockchain real estate, divided into blocks and incrementally released, are entangled but analyzed separately in this paper. 
In the first part of the paper, we consider what can be roughly termed the supply side, whereas, in the second part, we consider the demand side. 

Gas is central to our discussion. It is paid by blockchain users and plays an integral part in blockchain construction. Who receives the payment is blockchain-dependent. It can fund the maintenance of a blockchain and often helps block builders to prioritise transactions in the \emph{mempool}, i.e., the pool of pending transactions\footnote{These are transactions, which have been submitted to the \emph{mempool}, but have not yet been incorporated in the blockchain. }. 

 
 In the case of Ethereum\footnote{For an analysis of the link of network activity and gas fees for Ethereum see Koutmos\cite{k2023}, Liu {\it et al.}\cite{L2022}, Pierro {\it et al.} \cite{p2019}, Donmez {\it et al.}\cite{D2022} and Karaivanov {\it et al.}\cite{K2024}.}, the blockchain, through the Ethereum virtual machine (EVM), sustains the state and carries out instructions that update it. This could be a simple transfer of ether, the chain's native token, or a more complicated execution of a smart contract. It may be worthwhile to remind oneself that a transfer in isolation does not constitute a contract since a contract under English common law requires consideration.

 The next two sections on blockchain supply convert the multidimensional resource requirements into an updated gas cost formula. 
 Afterwards, in three sections on blockchain demand, the gas price is modelled as a fractional Ornstein-Uhlenbeck process following earlier work on weather derivatives, where this process turned out to be a sound choice. 

\noindent
The paper is rounded off with some general remarks. 
\section{Gas: The Supply side}
\noindent
The choice of gas price function determines largely what is included in the blockchain. From this, it derives its importance. The amount of gas fee users offer above the minimum allows a ranking based on the profitability of proposed transactions, mostly submitted to the generally observable \emph{mempool}. Block builders create from the transactions blocks with the largest economic benefit for them, which means maximizing the `priority fees' while staying within the constraint governing block size. This ignores MEVs or other possible side deals.  What maximises block builders' economic benefit is not self-evidently best for current and potential users and the wider set of nodes maintaining the network. 

Each group might be further subdivided. Users might have a variety of different latency requirements and cost sensitivities. Operational nodes have different storage and computational costs associated with transactions since some transactions are computationally more expensive, whereas others are dominated by storage or bandwidth requirements. The question of how to quantify and capture these different requirements naturally arises. 

A fully multidimensional cost would be a natural response since each resource has its own cost and limitations. This is always possible, but one less ambitious option is to retain simplicity for the multitude of users and shift as much as possible additional complexity to the smaller number of block builders. The toy model below will show how this can be done.

Weitzman in the paper ``Prices vs Quantities''\cite{We1974}
discussed the trade-off between implementing constraints on price and quantity. 
A detailed application to tokens, which we will not repeat,  can be found in  Buterin\cite{But2019}, while a more recent discussion is in Ndiaye\cite{Ni2024}.

We remark that both price and quantity can be employed to constrain usage. Network failure due to breaches of hardware-induced constraints on bandwidth, storage space, or computational capacity of nodes seems, from heuristic grounds, harder to reverse than mispricing, which might see businesses migrate to other blockchains but could be reversible through fine-tuning. The ultimate restraint on functionality is state growth\footnote{This is reflected in the current choice of gas pricing, where writing to a new `slot' in a state costs a multiple of writing to an existing `slot' in state.}, as  
 it slows down synchronisation time for new nodes, and in addition increases the cost of every future operation on the chain\footnote{Consequences of state growth are that the state no longer fits as easily into memory and updating takes longer.}. Constraining the number of transactions per block and, hence, state growth is essential.  For these reasons, only hard quantity constraints will be studied thereafter.
 As exemplified by the blog of Buterin\cite{But2024},
there has been interest in considering multidimensional gas prices.

The blog has parallels to a problem central banks were confronted with during the 2007-2008 financial crisis. 
Central banks aimed to expand the range of acceptable collateral at liquidity auctions. Instead of accepting only one type of collateral, they allowed two types, each associated with different credit risks. 
The question then arose: How should these auctions be organized? Can a single-round auction combine more complicated preferences?  

What was proposed by Klemperer\cite{klemp}\& \cite{kl2013} and adopted after discussions by the Bank of England was a single-round sealed bid auction, where bidders could bid for a fixed amount but offer a choice of collateral. 
Each type of collateral had an associated borrowing rate to reflect its different creditworthiness.
The Bank of England could then select for each bidder what collateral mix to accept for the provided liquidity.  

Related ideas were also developed by Milgrom\cite{Mi2009}, and linear programming provides the mathematical underpinning to solve such a constrained optimization problem.
Klemperer, who also worked jointly with Milgrom and others on the proposal,  provided a nice graphical way to determine an acceptable solution.
In the section on our toy model, we will develop an analogous framework for gas pricing. Although the overlap is not exact, our interest lies primarily in their approach's heuristics and graphical implementation rather than in the precise technical details.


In this analogy, the central bank corresponds to the block builders in the Ethereum network. The financial institutions requiring liquidity are analogous to the blockchain users who initiate transactions. Lastly, the collateral constraints financial institutions face can be likened to the hardware constraints block builders encounter.

In the Ethereum network, transaction fees are divided into the base fee, priority fee, and max fee; the \textbf{Base Fee} is a mandatory, algorithmically determined fee that adjusts based on network demand and is burned (destroyed) to reduce the total supply of Ether (ETH). \textbf{Priority Fee} is an optional fee set by users to incentivise miners to prioritize their transactions. It is paid directly to miners. \textbf{Max Fee} is the maximum amount a user is willing to pay per unit of gas. It includes both the base fee and the priority fee.

The effective fee per gas unit is the sum of the base fee and the priority fee. If the effective fee is less than the max fee, the user is refunded the difference. This mechanism tries to balance users' wishes to prioritise their transactions with a hard limit on total cost to protect them from excessive spending. 

\noindent
The data utilized in this analysis was extracted from the Dune \footnote{\url{dune.com}} database, employing a structured query to calculate daily median base fees and median priority fees over the past 360 days. The extraction process involved two primary components. Firstly, daily median base fees were computed by truncating the timestamps of block data to the nearest day and then calculating the median base fee per gas unit from the \texttt{ethereum.blocks} table. Secondly, daily median priority fees were determined by joining transaction data from the \texttt{ethereum.transactions} table with block data, again truncating the timestamps to the nearest day and calculating the median difference between the maximum fee per gas and the base fee per gas. To achieve the median values, these computations were performed using the \texttt{APPROX\_PERCENTILE} function. The final dataset was obtained by joining the daily median base fee and priority fee records on their respective dates. 
\begin{figure}[h!]
    \centering
    \includegraphics[width=\textwidth]{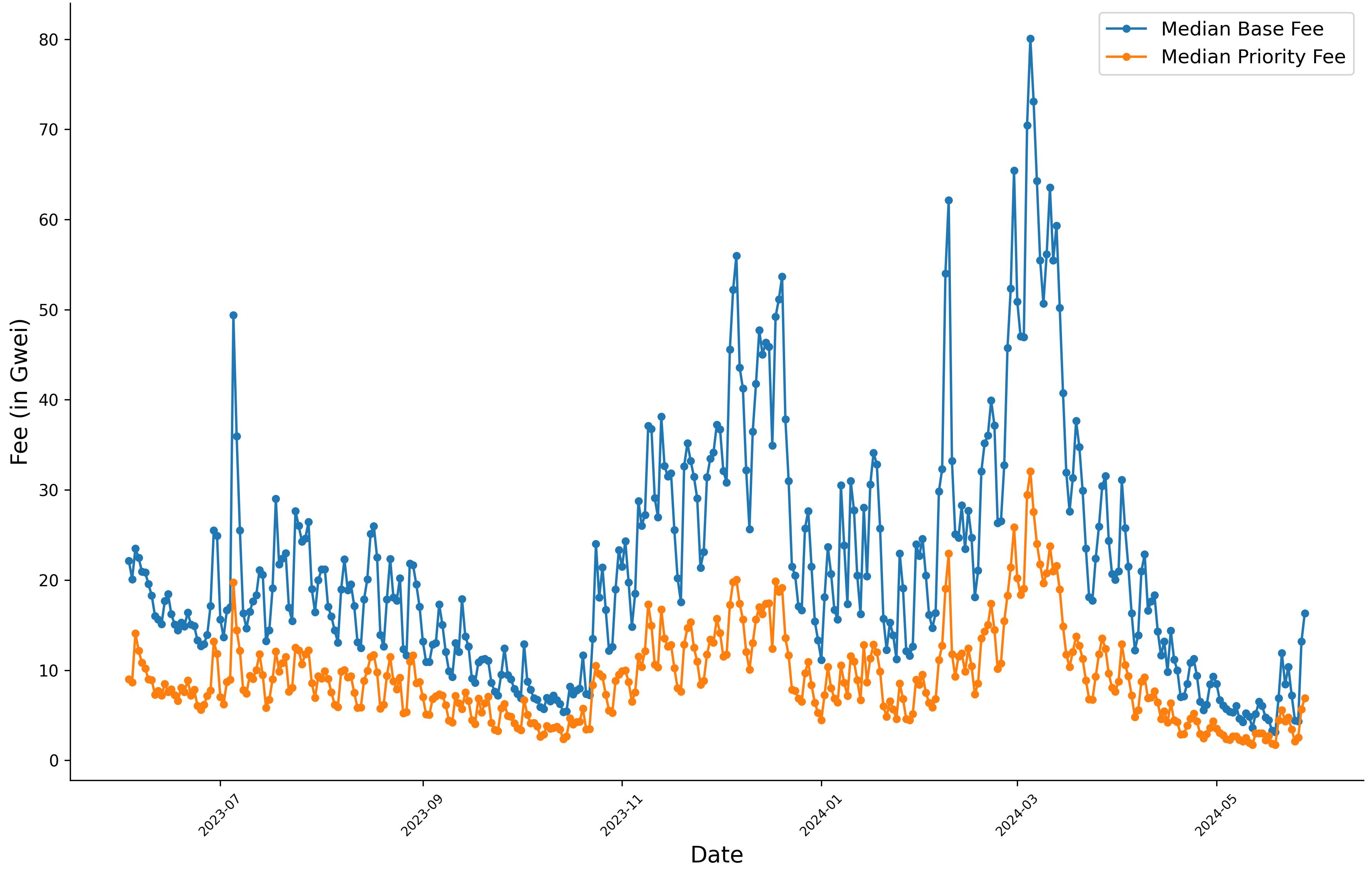}
    \caption{Daily Median Base Fee and Priority Fee Over Time. The data was extracted from the Dune database, where daily median base fees were computed from block data and daily median priority fees were computed from transaction data joined with block data. This analysis covers the past 360 days. 
    }
    \label{fig:daily_median_fees}
\end{figure}
In the next section we present a new gas model to improve the management of  resource constraints. 

\newpage
\section{Toy Gas Model For Incorporating Resource Constraints}
\noindent
Up to now we have described properties of the blockchain. In this section we take the next step, and 
 introduce a simplified supply model highlighting the multidimensional nature of the resources necessary for maintaining and expanding the blockchain. The core concept is as follows:
Transactions included in the blockchain can be split into operations. Operations require resources, which are constrained and can be transformed into gas. Gas can be converted into a fee or cost payable in ether. We use the terms fee and cost interchangeably. 


Let us define some additional terms. The gas cost paid for a transaction is a product of the amount of gas used and the cost per unit of gas. This can be further subdivided since gas splits in the model we consider into $M$ different resources. These are, for example, 
the amount of computation required or data stored, constituting the multidimensionality of gas. The availability of these resources adjusts on the time scale of months with technological progress and other external factors like electricity prices. Here, we assume that the limits per block for each of the $M$ resources are provided as an external input. Gas has the dimension of `units of gas' and is converted to ether by a quantity with the dimension `ether per units of gas'.

More about the different components that cost can be decomposed is described next. 
Each distinct operation involves the use of various resources. The `static gas' matrix captures this transformation. 
Besides the `static gas' there is a vector called `dynamic gas'. This vector will capture the individual resource constraints. It will be adjusted similarly to the `base fee' of \emph{EIP-1559}, except that not the total amount of gas of the past block but the 
gas associated with a particular resource provides the condition for adjusting the particular dynamic factor. 

The intuition is that \emph{EIP-1559} was about one overall constraint for gas implemented through the adjustment of the base fee, but here, we want to have a finer-grained constraint for individual resources. This is achieved by diversifying the conversion factor, i.e. adding `dynamic gas'. Overall congestion still will affect the `base fee', but fine-grained resource-specific congestion will affect through the `dynamic gas' coefficients the aggregate cost. This will allow multiple constraints without introducing multiple `base' or `priority fees'. The user experience will stay simple, while the, on average, more sophisticated blockbuilders are faced with marginally more computational effort. 

For the precise terms of the current framework of Ethereum, see the Yellow Paper\cite{wo2024}. 
The details of the applicable `Fee Schedule'  can be found in the paper's appendix G. It allows the transformation of operations\footnote{ There are about three dozen types of operations, including refunds, with fees ranging from one unit of gas  `for a JUMPDEST operation' to $ 32000$ units of gas `for a CREATE operation'. } of transactions or blocks into what is called a `gas limit'. This is the scalar quantity, which we will assume to be subdivided into different buckets, each with its own constraint.  To avoid confusion, we use, unlike the Yellow Paper, the term `fee' or `cost', as stated above, always in connection and priced in ether (denominated in ETH, Gwei or Wei, with $1 ETH\, =\, 10^9\,\, Gwei =10^{18} \,\, Wei$), and `resources'  are denominated in `gas'. 
The term `transformation' describes turning 'resources' underlying 'operations' into `gas', while `conversion' describes turning `gas' into cost in ether. This is to clarify the paper's idiosyncratic terminology, but it is otherwise of no importance.

Let's next turn the words into equations. For brevity, we sometimes use the  Einstein convention, which assumes repeated indices are summed over even without the summation symbol and requires consistent application of indices. Vectors are written with an arrow overhead, while matrices carry a double arrow. Both are written in bold and capital letters. Coefficients of vectors and matrices use the same letters but in lowercase. 
The description is given in three forms. In more symbolic notation, index notation, and visual arrays. The different representations are equivalent and are just added to benefit a diverse readership.

The Gas cost $C_t$ of block $t$,   where $t\in\{1,2,3,...\}$,  a scalar in our toy model,   is the product of multiple terms. It includes the scalar `base fee' and the vector `priority-fee', the $B_t\overrightarrow{\mathbf{I}}$\footnote{$\overrightarrow{\mathbf{1}}$ is the transposed of the vector $(1,1,...,1)$. } and vector $\overrightarrow{\mathbf{P}}_t$ respectively, where both terms are block $t$ dependent, and the priority-fee is in addition an $N$ dimensional vector.  
 The $i$-th transaction uses the amount $\overleftrightarrow{\mathbf{\Pi}}_t$, which is a $O_p$ times $N$ dimensional matrix\footnote{We define $O_p$ to be the number of distinct operations, as for example defined in the `Fee Schedule' of Appendix G of \cite{wo2024}. }, of resources.  
 The transformation of operations to resources is done by the matrix $\overleftrightarrow{\mathbf{G}}_t$, which is a $O_p$ times $M$ dimensional matrix\footnote{We define $M$ the number of distinct resources that each will have its own constraint.}. This matrix is modified by the `dynamic gas' transposed vector
 $\overleftarrow{\mathbf{\Lambda}}_t$, which is $M$ times $1$ dimensional\footnote{We use $\overleftarrow{\mathbf{A}}$ as the transposed of the vector $\overrightarrow{\mathbf{A}}$.}.
Combining the different elements produces the equation for the cost of a block of
 \begin{eqnarray}
   C_t = \overleftarrow{\mathbf{\Lambda}}_t\overleftrightarrow{\mathbf{G}}_t \overleftrightarrow{\mathbf{\Pi}}_t\big(B_t\overrightarrow{\mathbf{I}}+  \overrightarrow{\mathbf{P}}_t\big), \nonumber
\end{eqnarray}
which can be rewritten in terms of coefficients  
 \begin{eqnarray}
  c_t = \lambda_{k} (t) \, g_{kj} (t) \,\pi_{ji}(t)\big(B_t 1_i+  p_i(t)\big), \nonumber
\end{eqnarray}
and in terms of arrays 
\[  
c
=
\begin{pmatrix}
    \lambda_1      & \lambda_2 & \lambda_3 & \dots & \lambda_M \\
\end{pmatrix}
\begin{pmatrix}
    g_{11} & g_{12} & g_{13} & \dots  & g_{1O_p} \\
    g_{21} & g_{22} & g_{23} & \dots  & g_{2O_p} \\
    \vdots & \vdots & \vdots & \ddots & \vdots \\
    g_{M1} & g_{M2} & g_{M3} & \dots  & g_{MO_p}
\end{pmatrix}
\begin{pmatrix}
    \pi_{11} & \pi_{12} & \pi_{13} & \dots  & \pi_{O_p1} \\
    \pi_{21} & \pi_{22} & \pi_{23} & \dots  & \pi_{O_p2} \\
    \vdots & \vdots & \vdots & \ddots & \vdots \\
    \pi_{O_p1} & \pi_{O_p2} & \pi_{O_p3} & \dots  & \pi_{O_pN}
\end{pmatrix}
\begin{pmatrix}
     B_t  +  p_1  \\
    B_t  +  p_2 \\
    \vdots & \\
    B_t  +  p_N
\end{pmatrix}
. 
\] 
In the last representation of the cost equation, the $t$ dependence was dropped for notational brevity. 


If one wants to work with truly multidimensional gas such that users have to pay separate fees for each of the $M$ resources, then instead of the vector $B_t\overrightarrow{\mathbf{I}}+  \overrightarrow{\mathbf{P}}_t$ one would have to introduce a $N$ times $M$ dimensional matrix. The transposed vector  $\overleftarrow{\mathbf{\Lambda}}_t$ would be replaced by an $M$ times $M$  matrix $\overleftrightarrow{\mathbf{\Lambda}}_t$, as a general `dynamic gas' to allow cross-linked constraints and adaptation of the $M^2$ coefficients of $\overleftrightarrow{\mathbf{\Lambda}}_t$ with congestion.

Next, we introduce hard block-by-block constraints as a $M$ dimensional vector of the form $\overrightarrow{\mathbf{L}}$ with coefficients $l_{ k}$, we further assume that the sum $\sum_{k=1}^{M}l_{ k}$ corresponds to the current number of target gas units per block\footnote{The current target gas is $15\,\times\,10^6$ units of gas per block, whilst the hard cap gas limit is $30\,\times\,10^6$ units of gas per block. It is worth noting that this is not a `hard-coded cap' but an arrangement agreed on by the validators en masse - see ethereum research \url{https://ethresear.ch/t/on-block-sizes-gas-limits-and-scalability/18444}}. These values for the resource constraints are set externally and updated infrequently with technological advances and other changes influencing the EVM.
Each of the `dynamic gas' coefficients $\lambda_{ k}$ is deemed to start at one and ideally should mean revert around this value. How is this achieved? 
For a fixed $\hat{k}$, if the block's $k$-resource  $\sum_{i=1}^N g_{\hat{k}i}(t)$ is bigger than $l_{\hat{k}}$,  then the value of  $\lambda_{\hat{k}}$ applicable to the block $t+1$ will be increased and, if it falls below the $l_{\hat{k}}$, then the value will be proportionally decreased using a formula along the line of \emph{EIP-1559}\footnote{The EIP-1559 base fee adjustment formula for overall congestion is given in terms of `gas used', 'target gas', current 'base fee', updated 'base fee' and some constants. 
For details, see \cite{wo2024} and the many papers written to analyze the proposal.}
, but dependent on `relative congestion' in terms of the $\hat{k}$-th resource instead of `absolute congestion'.  


\vspace{.2cm}

In the next paragraphs, we will show how block builders can fish out of the \emph{mempool} the most profitable transactions. As before, MEVs or other side deals are ignored. 
There is a simple graphical representation of transactions that enables easy comparison. 
Each transaction corresponds to a line that intercepts the resources axes.  The points of intersection are chosen such that the `priority fee' is exclusively associated with just one resource consumed by the transaction. 
This leads to a hyperplane of $M-1$ dimensions for each of the $M$ constraints.
If no cost of a particular kind is associated with a transaction, the line in the two-dimensional case is parallel to that axis. 


In this paragraph, we delve deeper into the graphical representation. 
Assume the two-dimensional resources case and set the dynamic gas coefficients to one.  Let's further assume that the transaction under consideration requires `$x$' units of gas for `storage' and `$y$' units of gas for `computation'. The priority fee is set to `$z$' Gwei per unit of gas.  The total amount of gas is given by $x+y$; the total priority cost is $(x+y)z$. 
The equation of the relevant line is given by $x\, \mathbf{S}+ y \, \mathbf{C}=(x+y)z$, where $\mathbf{S}$ represents the `storage resource' and $\mathbf{C}$ represents the `computational resource'. The line shows how the total priority fee can be attributed to the two resources in different ways. 
Figure 2 (a) depicts the case with `$x$' equal to one , `$y$' equal to two,
and `$z$' equal to $3/4$, since the line segment fits the equation $\mathbf{S}  + \frac{1}{3} \mathbf{C} = 1 $.

This representation has the advantage that different transactions cannot only be compared, but furthermore, transactions are represented by identical lines as long as they are related by a scale factor, i.e. the same line represents the triplets $(x,y,z)$ and $(kx,ky,z)$. From the perspective of block builders the identification of lines that have the same `priority fee' per unit of gas and the same ratio of resource requirements is reasonable since incorporating $k$-times the smaller transaction $(x,y,z)$ into a block is equivalent to the inclusion of one big transaction
of the form $(kx,ky,z)$ in terms of `base' and `priority fee' as well as resource requirements\footnote{Blockchain users view this differently since combining transactions can be beneficial, e.g. there is a base cost for each transaction of currently $21,000$ Gas. Access lists can also `warm up' storage for a transaction prior to accessing it (see EIP-2930: Optional access lists). 
To clarify and justify the scaling assumption for block builders, the resources charged and the resources consumed by block builders must match. An exception exists as a block approaches a hard limit for one of the resources; a block builder could prefer smaller transactions for added flexibility.}




\begin{figure}[h!]
  \centering
  \begin{subfigure}{0.48\textwidth}
    \includegraphics[width=\linewidth]{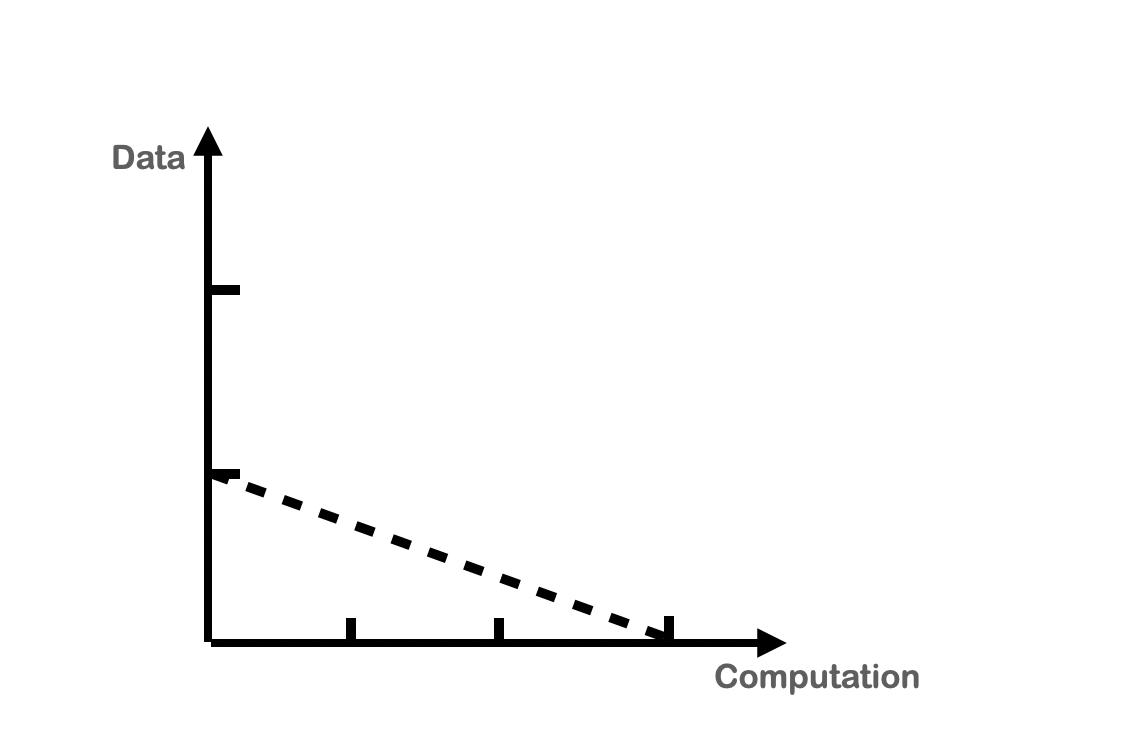}
    \caption{ }
    \label{fig:a1}
  \end{subfigure}
  \hfill 
  \begin{subfigure}{0.48\textwidth}
    \includegraphics[width=\linewidth]{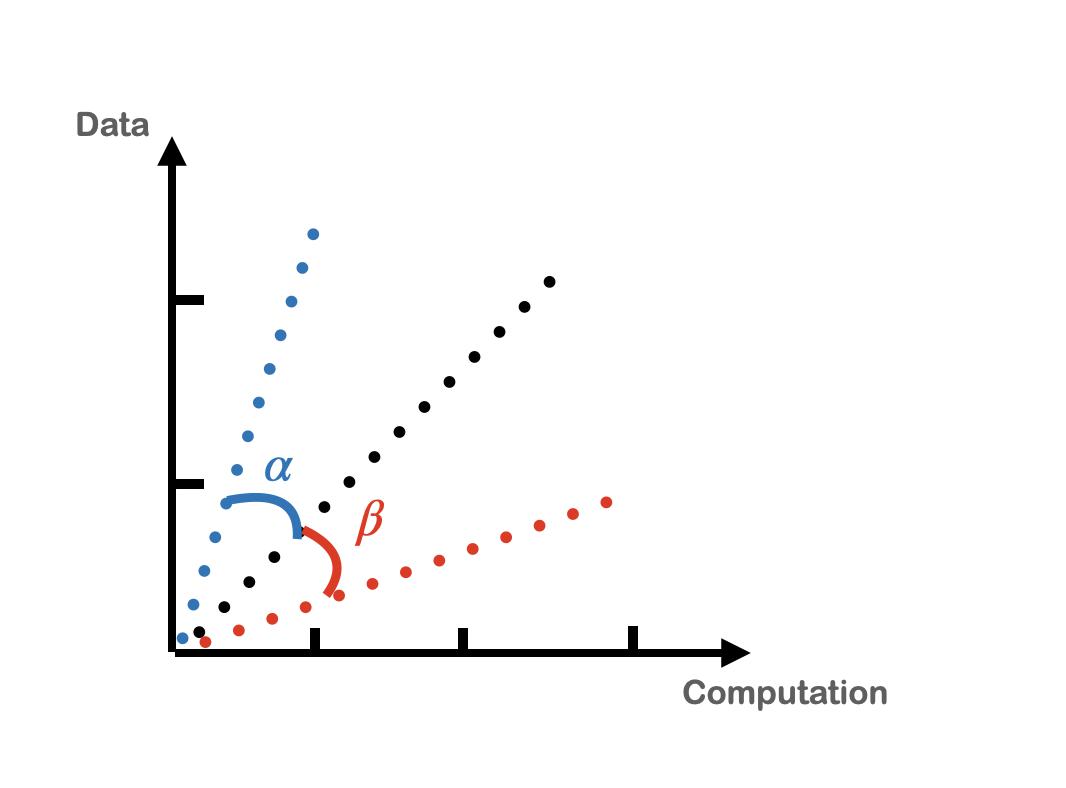}
   \caption{}
    \label{fig:b}
  \end{subfigure}
  \caption{ (a) Line segment intercepts the `storage' axis, where the normalised priority fee is all assigned to `storage', and intercepts the `computation' axis, where the normalised priority fee is all assigned to `computation'. 
  (b) The red line corresponds to gas weighted resources vector, i.e. the coefficients for each of the $M$ resources are equal to the amount of `static gas' associated. In contrast, the blue line is a reflection of the red line on the vector $(1,1,...,1) $ and is the direction in which the optimal search is carried out.}
\end{figure}
\begin{figure}[h!]
  \centering
  \begin{subfigure}{0.48\textwidth}
   \includegraphics[width=\linewidth]{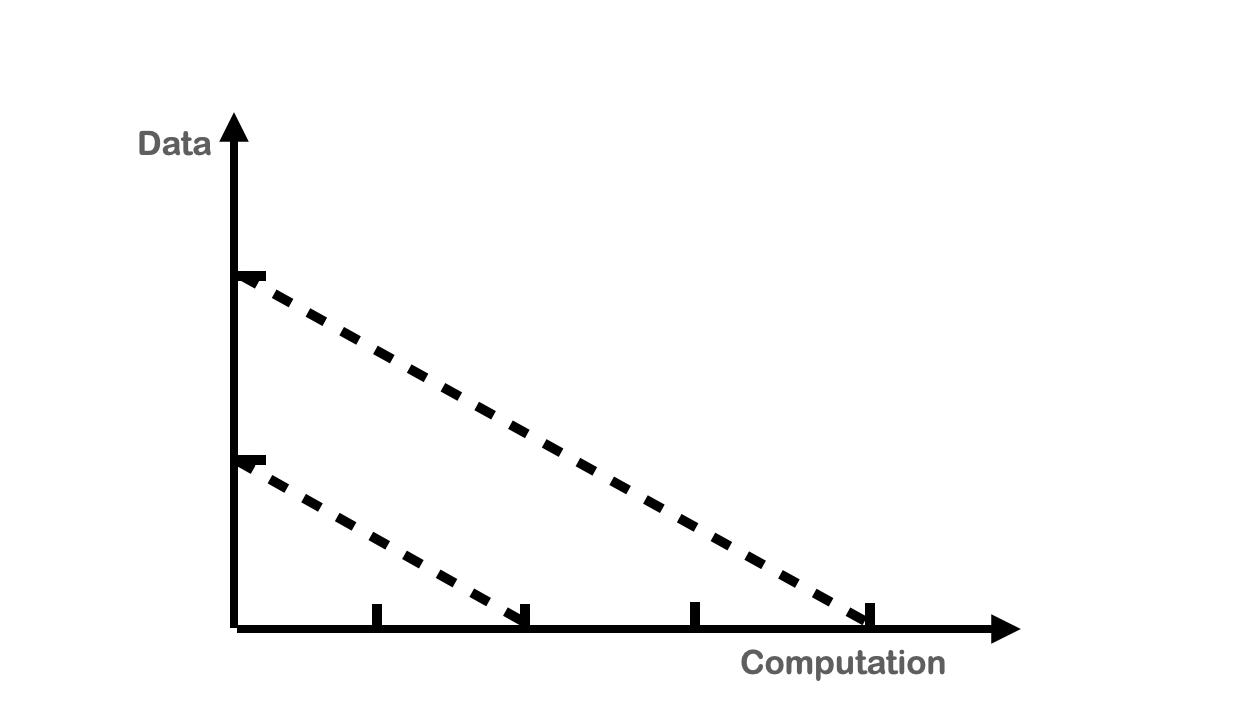}
    \caption{}
    \label{fig:c}
  \end{subfigure}
  \hfill 
  \begin{subfigure}{0.48\textwidth}
      \includegraphics[width=\linewidth]{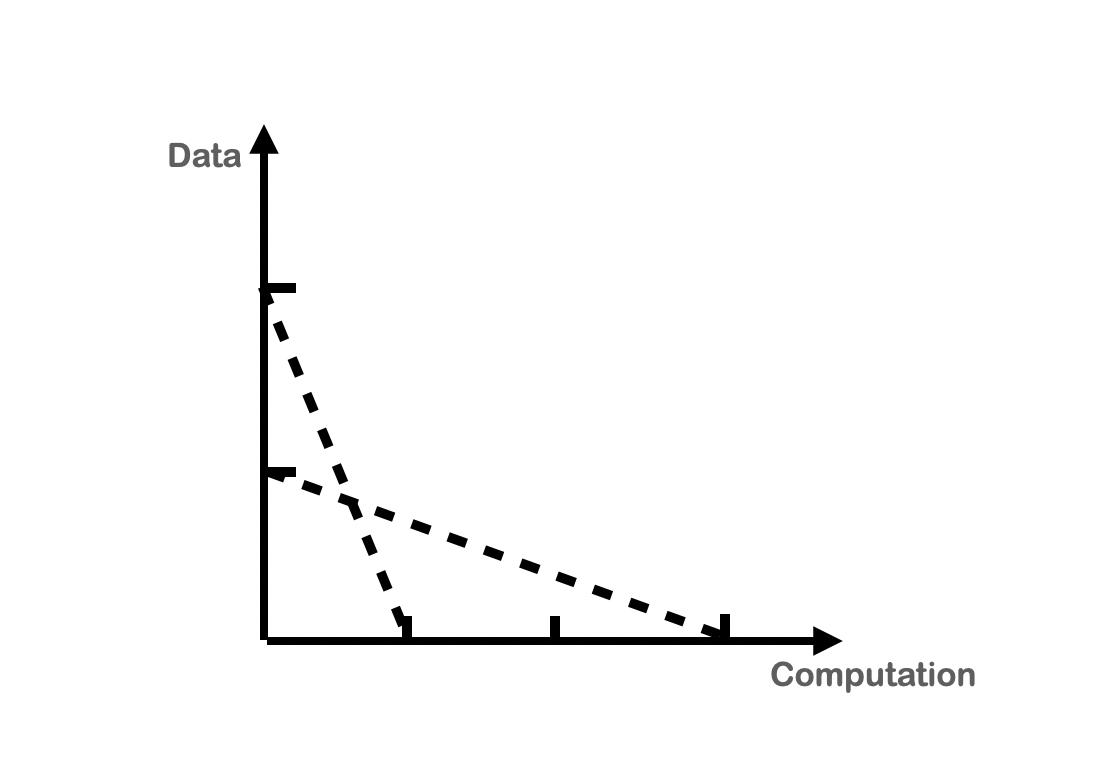}
    \caption{}
    \label{fig:d}
  \end{subfigure}
  \caption{(a) Two line segments, each representing a type of transaction, with the one closer to the origin always being more attractive independent of the relative resource constraints.
  (b) Two intersecting line segments. The preferred transaction depends on the relative resource constraints.  }
\end{figure}


\begin{figure}[h!]
  \centering
   \begin{subfigure}{0.48\textwidth}
    \includegraphics[width=\linewidth]{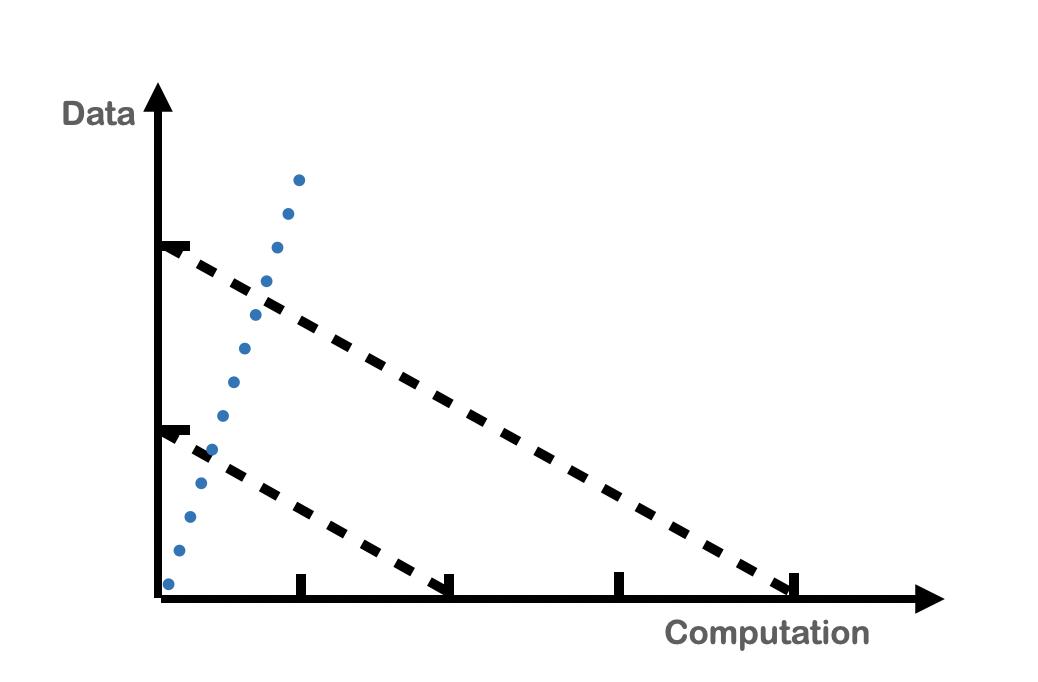}
    \caption{}
    \label{fig:d}
  \end{subfigure}
  \hfill 
 \begin{subfigure}{0.48\textwidth}
    \includegraphics[width=\linewidth]{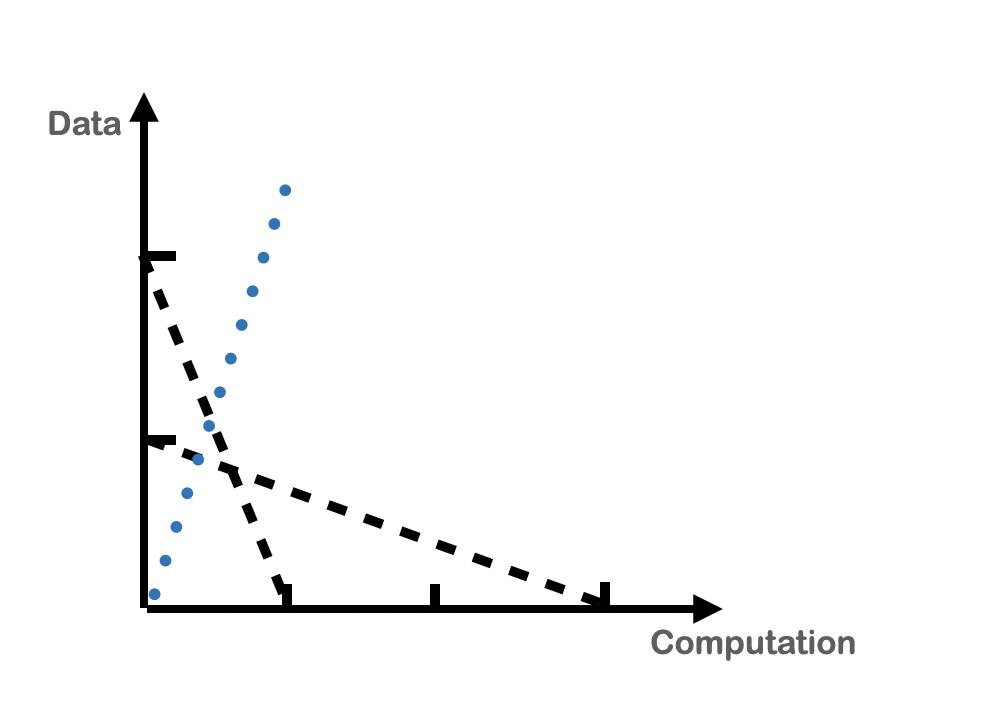}
    \caption{}
    \label{fig:c}
  \end{subfigure}
  \caption{(a) The blue vector, independent of direction, always intersects one line first. 
  (b) The line first intersected by the blue vector depends on the direction of the blue vector. }
\end{figure}
Two further cases are considered in the two-dimensional criteria case to acquire an intuition.
We compare two transactions with corresponding lines that intersect and which don't intersect. 
Suppose the two lines do not intersect (see Figure 4 (a)). In that case, the transaction associated with the line closer to the origin is preferred independent of the constraints, assuming one is sufficiently far from any constraint boundary. 
If the two lines intersect (see Figure 4 (b)), then the choice depends on the relative size of the coefficients of the constraint vector  $\overrightarrow{\mathbf{L}}$.  
The relevant direction of the vector  that intersects the lines associated with the transactions is given in the two dimensional case by $(L_2,L_1)$.
As an example, if $\overrightarrow{\mathbf{L}} = (2,1)$ then the direction of the blue test vector shown in blue (for Figure 4 (a) $\&$ (b)) is $(1,2)$ starting from the origin. 
In the general case, with an original weight induced vector of $\overrightarrow{\mathbf{L}}$, relating the different constraints, being reflected on the $(1,1,...,1)$ vector will produce the new vector $\overrightarrow{\mathbf{T}}$. The equations determining the new vector are: 
$\overrightarrow{\mathbf{T}}=  \cos{\alpha} \,\overrightarrow{\mathbf{E}} - \overrightarrow{\mathbf{E}}_{\perp}$,
with $\overrightarrow{\mathbf{E}}_{\perp}=\overrightarrow{\mathbf{L}}-\cos{\alpha} \overrightarrow{\mathbf{E}}$,
and 
$\overrightarrow{\mathbf{T}} 
=2\, \cos{\alpha} \,\overrightarrow{\mathbf{E}} -\overrightarrow{\mathbf{L}}$,
with $\cos{\alpha}:=\overrightarrow{\mathbf{L}} \overrightarrow{\mathbf{E}}/ \big(\|\overrightarrow{\mathbf{L}}\|\,\|\overrightarrow{\mathbf{E}}\| \big) $


This toy model goes incrementally beyond \emph{EIP-1559} since it ``start[s] with a base fee amount which is adjusted up and down by the protocol based on how congested the network is" (taken from the EIP-1559 proposal), but instead of having one quantity one has $M+1$ quantities. 
The first $M$ limits adjust relatively slowly with technological advances, and the last constraint is the sum of the others.
The decomposition and number of constraints will also evolve over longer time scales. 

Which function is suitable for the modification of the `dynamical factor' $\overleftarrow{\mathbf{\Lambda}}_t$,  if one wants to go beyond the \emph{EIP-1559} adjustment mechanism applied to relative constraints, depends on the drift and volatility encountered in the demand for resources.  One can use the sophisticated method described in \cite{Pe2020} to discern a drift in noisy data. Another simpler way is to smooth the data to generate stability but at a cost in reaction time. 
In general, if volatility is high, a faster response might be needed, but a correct response, which avoids overreaction, might be harder since observed volatility can be due to real market changes, i.e., different contracts getting popular or the irregular arrival of transactions. 

Hardware questions that flow into the choice of resources to be considered and constraints to implement need to be discussed in a different setting geared towards engineering.  We assume that the  changing nature of hardware, software and general infrastructure 
can, at any point in time, be translated into a well-defined set of constraints.



 General transaction fee mechanism design, and \emph{EIP-1559} in particular, has been scrutinised by  Chung {\it et al.}\cite{ch2021}, Roughgarden\cite{Ro2021}   and others.   Some see it as an unalloyed success; others are more circumspect. See for the description of some challenges (also applicable to the Bitcoin protocol)  Kiayias {\it et al.}\cite{ki2023}. As an alternative, they propose as a solution for congestion and to enhance `egalitarianism' a tiered mechanism, where users can choose the urgency and 
capacity is set aside for cheaper low latency demand. 
This prevents the withdrawal of lower-value transactions from a chain during high throughput. The authors compare their proposal to a service-dependent price, where a high price results in speedy execution, but a low price is prone to delays. 
The authors draw an analogy between this outcome and a multi-speed highway. To conduct experiments on queuing times, one can augment the number of channels and adjust their width. This comparison also evokes the image of a theme park with VIP lanes, where operators could delay standard ticket holders to provide preferential treatment to VIPs, ostensibly maximizing their perceived value.
A problem remains: if endemic demand outstrips processing ability, then high-vaulting `inclusivity' intentions will not prevent the failure of some transactions to be processed.
If demand-supply imbalances are temporary, then their scheme or similar schemes proposed by others might be an interesting way to allocate capacity. 
One could derive a possible inducement for deleting dormant smart contracts, which take up space on the blockchain, from what has been observed for the exercise times of options - American options with flexible exercise time versus European options with fixed exercise time. 
Under standard conditions ignoring dividends\footnote{Also ignoring other extraneous factors like stock borrow.} it is best to exercise options at the latest possible time. This could be similar to applying for refunds and deletions of smart contracts under the current scheme. To encourage deletions, one could add a discount function,  
or with each technological change that increases \& cheapens storage and therefore devalues stored data\footnote{This is similar to dividends in American options, which, under the right condition, induce early exercise.}, one could reduce the returned
amount.  

Next, some examples of the diverse blockchain network fee structures currently in fashion.
On many blockchains, transaction size and demand determine fees.
Bitcoin, the most prominent token, employs a transaction fee model. Fees are determined by the transaction size in bytes and current network demand. Users are incentivized to offer high enough fees so that miners select their transactions as they strive to maximize revenue. 
In contrast to Bitcoin,  IOTA employs a highly unusual feeless transaction model using its Tangle architecture, a Directed Acyclic Graph (DAG). Each transaction in the IOTA network confirms two previous transactions, thus eliminating the need for miners and transaction fees. Other blockchains allocate resources in a dramatically different way.
EOS, for example, adopts a resource allocation model in which users stake tokens to access network resources such as CPU and RAM. Instead of paying per transaction, users can perform transactions ``for free" within the limits of their staked resources. The above examples hint at the great variety on offer.
\noindent 
 In the next two sections, we switch to the demand side and consider a pre-determined gas price process.  
 \section{Gas Price Dynamics: An Introduction to the Demand Side}
\noindent
 Gas prices, as observed for various tokens, show persistence and mean-revert. Fractional Ornstein-Uhlenbeck processes possess both these properties and seem a good candidate as a base process. This approach has also proven to be popular for temperature modelling in the field of weather derivatives. Similar to gas prices across blocks, a direct investment in temperature across different time periods is not possible. Instead, gas prices can be observed and then incorporated into derivatives contracts. 

 The relationship between Brownian motion and fractional Brownian motion is similar to the relationship between the Ornstein-Uhlenbeck and the fractional  Ornstein-Uhlenbeck process. In both cases, the fractional process represents a generalisation by adding one parameter linked to the autocorrelation.  
 
Various simplifying assumptions are applied when modelling gas fees. In each block, gas fees per unit of gas, in terms of `base' and `priority fees', vary across included transactions. We could model the outliers, e.g. the maximum or the minimum fee per block or some average. Under most conditions, due to a lower limit to the gas fee, the volatility of the maximal gas fee should be higher than the volatility of the mean, median or minimal gas fee, even if very easily counter-examples can be constructed. Here, we take the median as the gas fee to be modelled. Higher-order moments of the gas are also not considered but are necessary for a comprehensive description of the problem.

\begin{figure}[ht]
    \centering
    \includegraphics[width=\textwidth]{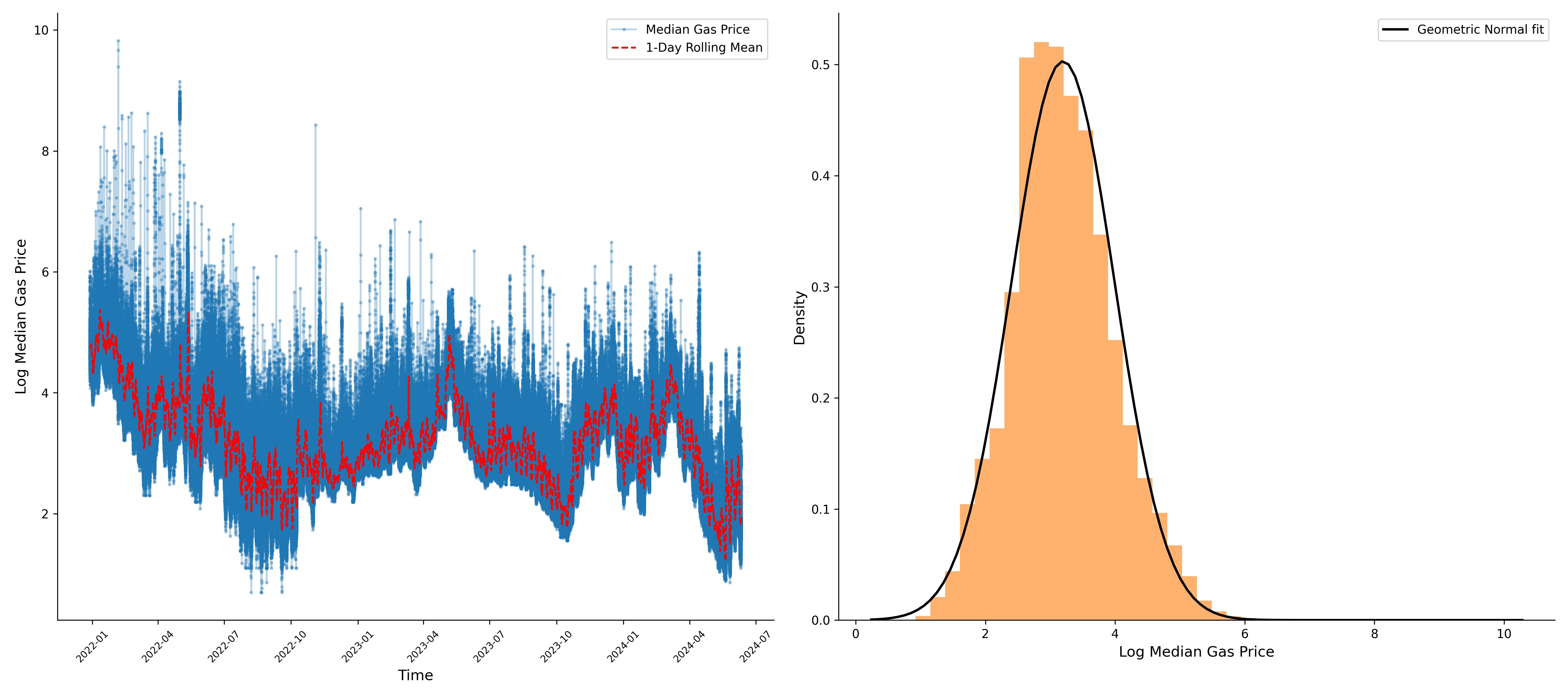}
    \caption{Motivation for a Geometric Fractional Ornstein-Uhlenbeck Process. The left plot shows the median gas price over time, including minute-level data and a 1-day rolling mean to illustrate mean reversion. The right plot displays a histogram of the median gas prices with a log-normal fit, highlighting the goodness-of-fit metrics: $\mu = \mu_{value}$, $\sigma = \sigma_{value}$. The data is sourced from Dune.com and represents the  minute-level data for approximately 900 days. 
    }\label{fig:median_gas_price_analysis} 
\end{figure}

The modelling raises several questions that will be discussed near the end of the paper.
Is it, for example, possible to simplify planning by creating besides a `spot market' for position in the blockchain also a `forward market'? In commodities, `forward markets' are useful for producers and users to hedge risk.

\section{A Gas Model: Fractional Ornstein-Uhlenbeck Process}
\noindent
In this section, Fractional Brownian motion (fBM), a generalisation of Brownian motion characterised by the Hurst exponent $H$,  is formally introduced. It forms the basis of a mean-reverting Fractional Ornstein-Uhlenbeck process that describes the temporal evolution of the gas price. 
The fBM is described in detail in the book by Biagini {\it et al.}\cite{Bi2002}.  The mean-reverting application to weather derivatives can be found in Brody {\it et al.}\cite{bsz2002}, and their economic notion is used throughout the section.
Fractional Brownian motion $W^H$, parameterised by the Hurst exponent $H\in (0,1)$, and with $H=1/2$ corresponding to the conventional Brownian motion,
is a Gaussian stochastic process  defined on  $(\Omega,  \mathcal{F}, \mathbb{P}^H ) $.
The sample path of the process $W^H$ are continuous  with $W^H_0 = 0$, and for $t\geq 0$
 \begin{eqnarray}
 \EX [ W^H_t W^H_s]=(t^{2H}  +s^{2H}-|t-s|^{2H}).
 \nonumber
 \end{eqnarray}
 For $H> 1/2$, the correlation between the increments is positive, while for $H< 1/2$, the correlation between the increments is negative.
$W^H$ is ‘self-affine’, i.e. $W^H_{\alpha t}$ 
has the same distribution as $\alpha W^H_t$ for every positive $\alpha$.

Some background about the fBM is given next.
The Hurst exponent 
takes its name from the hydrologist H.E. Hurst, who studied the time series of water levels of the Nile in the middle of the last century. He noticed long-range dependencies and scaling behaviour. This may not be surprising since cumulative precipitation in a river's catchment area drives water levels downstream.  
Some decades later, Mandelbrot named the parameter $H$ in the fBM in honour of  Hurst.
The gas cost $X_t$ at time $t$ is now defined
as the following process
 \begin{eqnarray}
 dX_t =\kappa (\theta - X_t)
 dt+\sigma dW_t^H, \,\,\, X_0=x. \nonumber
 \end{eqnarray}
The above equation can be solved under some simplifying assumptions for the parameters as shown in Duncan {\it et al.}\cite{duncan2000} and developed in Hu and Oksendal\cite{HO1999}. If 
$\kappa, \theta$ and $\sigma$ are bounded deterministic function, allowing us to write 
\begin{eqnarray}
 K_t = exp \Big( - \int_0^t \kappa_s ds \Big), \nonumber
 \end{eqnarray}
 then
 \begin{eqnarray}
X_t = x K_t + K_t \int_0^t \kappa_s \theta_s K_s^{-1} ds+
K_t \int_0^t   \sigma_s K_s^{-1} dW_s^H , \nonumber
  \end{eqnarray}
and further for $t\geq 0$ $X_t$ is a normal random variable with the mean $m_t^X= \EX [ X_t]$,
 \begin{eqnarray}
 m_t^X=x K_t + K_t \int^t_0 \kappa_s \theta_s K_s^{-1} ds,\nonumber
  \end{eqnarray}
and with the variance  $V_t^X= var [ X_t]$, 
 \begin{eqnarray}
 V_t^X=K_t^2 \int^t_0 \int^t_0\sigma_u \sigma_s  K_u^{-1}K_s^{-1}\phi (u,s)  du ds\nonumber
  \end{eqnarray}
  with $\phi(u,s) = H(2H-1)  | u-s|^{2H-2}$.
  Material on fraction Brownian motion, like the above results, with application to weather derivatives can be found in Brody {\it et al.}\cite{bsz2002}.
  
  We assume for simplicity that gas prices are autocorrelated and form clusters of high and low prices. This can be achieved by a fractional Ornstein-Uhlenbeck model, where the gas price
meanders around a pre-determined value, which can be cyclical, like some seasonal commodity prices or temperature. 
The gas price $g_t$ (see Figure 6) is described by
\begin{eqnarray}
d g_t = \kappa_t ( \mu_t- g_t) dt + \sigma_t d W_t^H\nonumber
\end{eqnarray}
with 
$\mu_t$ the mean reversion level, $\kappa$ mean reversion speed or mean reversion rate,
and $\sigma$ the volatility.

\begin{figure}[ht]
    \centering
    \includegraphics[width=\textwidth]{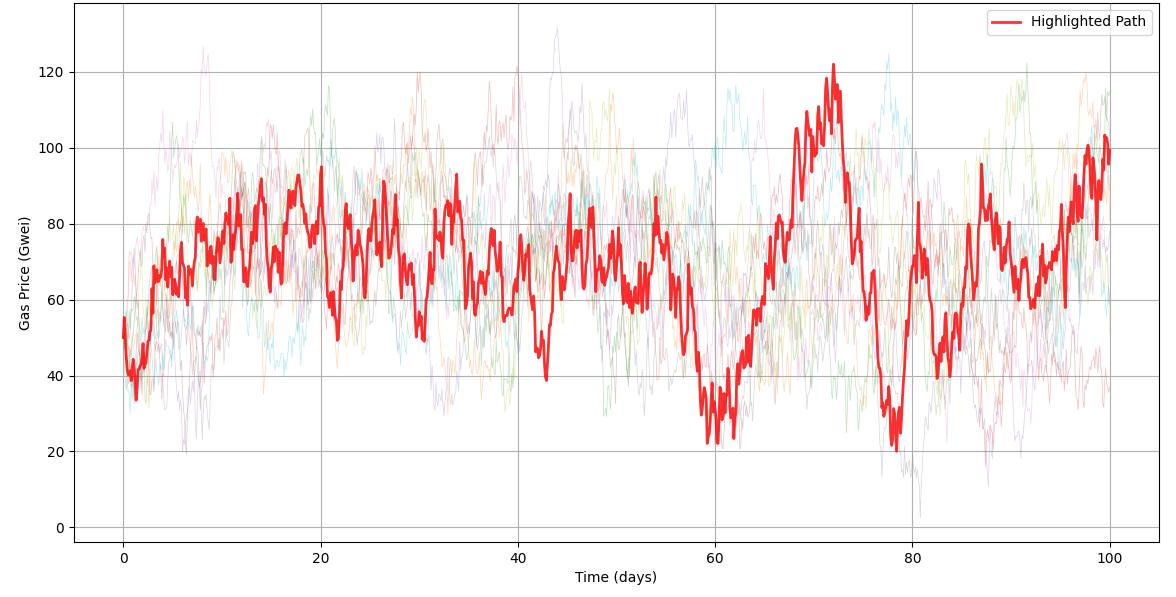}
    \caption{ Simulated Ethereum Gas Price Path Trajectories as a Fractional Ornstein-Uhlenbeck Process }\label{fig:amedian_gas_price_analysis}
\end{figure}

 
As an aside, the gas cost of a single block can be defined in multiple ways. Is it the highest or the lowest cost transaction accepted by the block builder, or is it the median or average cost? How are the volatilities of these different quantities related?  Except under 
artificial circumstances, which can be easily constructed\footnote{The maximal cost can be held constant, while the minimal, average, median cost can vary.} but are not replicated in the data; the volatility of the different percentiles of cost can be ordered with the lowest percentile having the lowest volatility since the cost is bounded below by zero.

\section{Gas Price Derivatives}
\noindent
This section explores different forms of gas fee derivatives. The fractional Ornstein-Uhlenbeck process developed in the previous section is employed as the underlying price process. 
European and other option prices for assets following the Ornstein-Uhlenbeck price process, i.e. $H=1/2$, 
can be found in \cite{Li2006, Heston2000, Hillard1978}. The generalisation to any $H\in (0,1)$ was considered for weather derivatives in 
Brody {\it et al.}\cite{bsz2002}.
Weather derivatives often involve the number of days a local  temperature falls below or rise above a fixed level $K$, which are called {\it heating} or {\it cooling} degree days. The associated payouts, respectively, for a specific day $t$ are
\begin{align*}
 ( K- X_t)^+,\nonumber \\
 (  X_t- K)^+.\nonumber
\end{align*}
The same quantity is of interest also for gas derivatives, and the abstract evaluation formula  of the the put and call version of the derivatives can be written in the form
\begin{align*}
P_t =\EX \Big[  e^{- \delta (T-t)} \int_{T-S}^T  ( K -X_t )^+ ds\vert \mathcal{F}_t^X\Big],\nonumber\\
C_t =\EX \Big[  e^{- \delta (T-t)} \int_{T-S}^T  ( X_t-K )^+ ds\vert \mathcal{F}_t^X\Big].\nonumber
\end{align*}
In addition, one can define a modified derivative with an additional strike price $L$
\begin{align*}
\EX \Big[  e^{- \delta (T-t)}\Big(\int_{T-S}^T \big( ( K -X_t )^+- L\big)^+ ds\vert\mathcal{F}_t^X\Big)^+\Big],\nonumber
\end{align*}
in this or in other permutations. 
All these derivatives have been priced, and the result can be found in section 4 of \cite{bsz2002}. 

An example of call option values for different initial gas prices and maturities is given in Figure 7.
Since the underlying is in the gas case not directly tradable, i.e. the price of the derivatives cannot be simply replicated or the product directly hedged.
Next comes a short conclusion rounding of the paper.

\begin{figure}[ht]
    \centering
    \includegraphics[width=\textwidth]{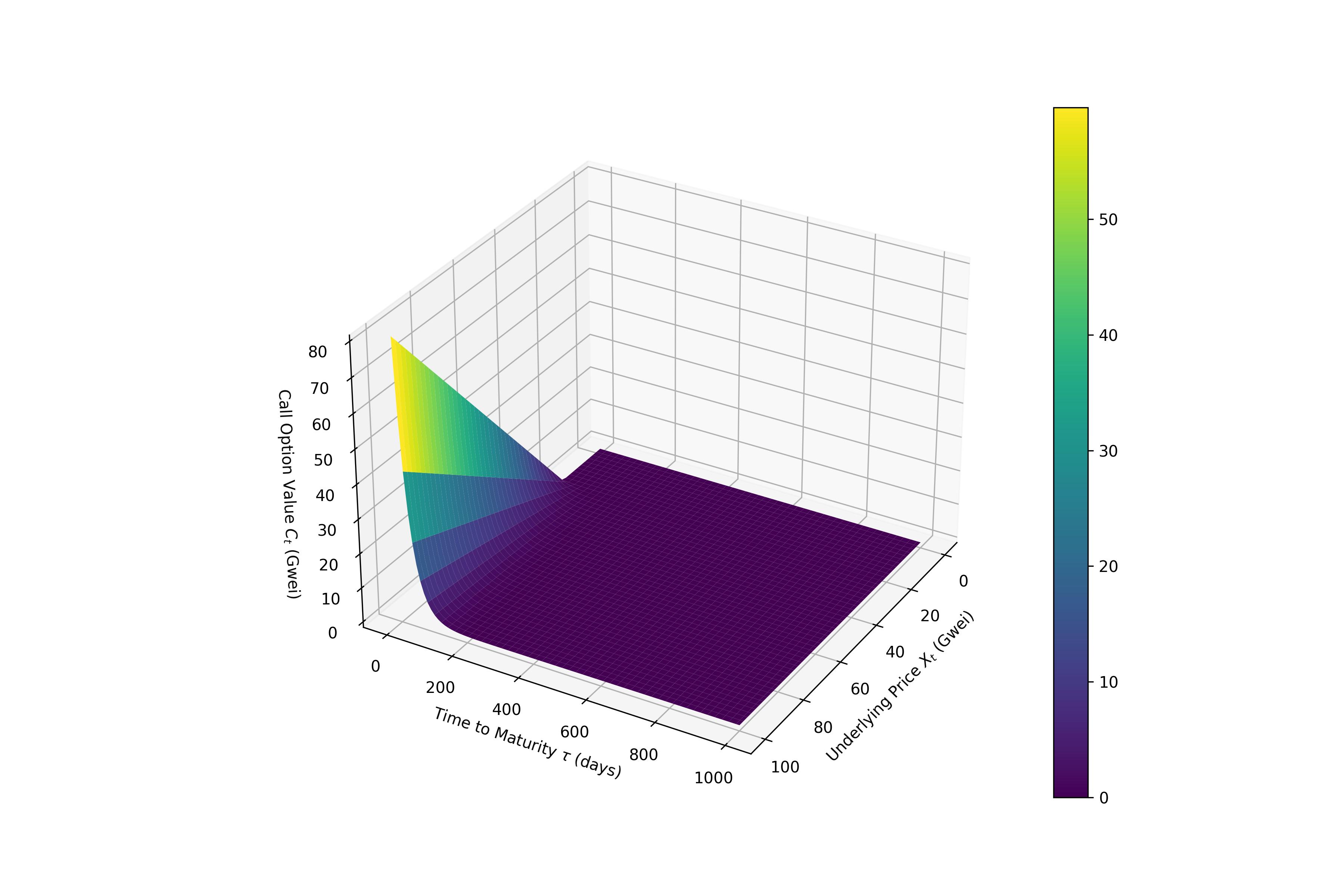}
    \caption{  Call Option Values for different times to expiry and underlying prices (Ethereum Gas), where the underlying follows a Fractional Ornstein-Uhlenbeck Process.  }\label{fig:amedian_gas_price_analysis}
\end{figure}

\section{Conclusion}
\noindent
Gas is not only a determining factor for users in their choice of blockchain but also for token holders, stakers and node operators. Each group has a different perspective.
On the supply side, node operators have to worry about shifting technological constraints and the costs involved. On the demand side, chains compete fiercely in Darwinian fashion. 

In the toy model introduced in section two and developed in section three, we showed how multiple constraints can be incorporated into the gas fee simply. Adopting multidimensional gas constraints will better align a blockchain with technological realities.

The subsequent sections focus on the demand side and how the fractional Ornstein-Uhlenbeck price process can act as a base for evaluating derivatives, similar to developments in weather derivatives. 
Temperature is a key variable to understand the weather but not directly tradable, unlike company ownership or obligations represented as stock or bond  certificates\footnote{Commodities are a directly investable quantity, but to understand their price processes, one is required to include storage cost and convenience yield\cite{kaldor}. }.
The same can be said for the gas fee. The magnitude of the gas fee has economic consequences, but there is no gas fee certificate that is directly investable or transferable between blocks. Consequently, the approximate continuous time series of gas fees does not have to follow a stochastic process that, under a change of measure, can turn into a martingale to avoid arbitrage.
Instead, gas fees on the Ethereum blockchain, like temperature, show identifiable patterns and have properties of a mean-reverting fractional Ornstein-Uhlenbeck process. 
This is theoretically appealing  and retains its plausibility after analyzing  the data. 
One of the interesting features of this model is that if transactions are not time-sensitive, execution delays can be traded off for cost improvements. 
 From the instantaneous drift of the fractional Ornstein-Uhlenbeck process, the cost improvement versus time delay can be read off.  This helps users to balance cost against time. 
Optimal investment strategies for an asset following the related Ornstein-Uhlenbeck process can be found in \cite{lv1, lv2}. 

 This raises the question of how users can hedge against potential gas fee spikes or the related questions of how one hedges an option when the underlying has a drift. How does one hedge gas cost since one cannot hand over block openings from block to block?
 How does one deal with blocks containing a whole host of transactions, not just one? 
These more practical questions will be addressed in a separate report.  

What are some of the limitations of the models proposed and related implementation challenges? The models were derived from first principles and justified by stylized facts. Only time will tell if a multi-dimensional gas fee model, as described in the paper, will receive a positive reception.  The same applies to gas fee forwards. The willingness of market participants to use such derivatives depends on the future cost and volatility of gas fees, which is dependent on the development of the Ethereum ecosystem. Both the future demand for blockchain space as well as protocol changes are inherently unknowable. 
Implementation challenges abound due to above stated uncertainties. 

One possible extension of the research, suggested to us by a referee, is to look at basis risk between expected spot and forward gas fees. Basis risk is the mismatch between the asset or liability one wants to hedge and the contract one uses as a hedge. The cause for the imperfect substitution can be market forces related to supply and demand imbalances or discrepancies in the contract definitions. Initially,  market makers are likely to have limited risk tolerance for gas fee forward contracts and will provide limited liquidity. As a consequence, imbalances in hedging demand \& supply between block chain users and stakers are likely to lead to mispricings and, consequently, basis risk. 

Another possible extension of the research is to develop not just the various derivatives proposed and priced in the paper but also forward contracts for gas prices at a time point or averaged over a time period in the future. Deployers and users of smart contracts would then be able to gain confidence that costs are predictable and hedgable. A short position in forward contracts has intrinsic attractiveness to stakers and block builders who are long gas fees, but could limit their downside by locking in the forward price. 
This raises other questions.
Could one pre-order block capacity through a forward market to reduce volatility? This benefits both sellers and buyers who wish to obtain fixed costs.  These commitments could be enshrined for future reference in current blocks.

In general, assuming the fractional Ornstein-Uhlenbeck process is well suited to modelling gas prices, the question arises: Can one use the predictive nature inherent in the process to improve the block creation protocol to increase efficiency?
This will be explored elsewhere. 

Another hotly debated topic is how to rein in Maximum Extractable Value (MEV). Reducing MEV might be in the purview of smart contract developers, who should have incentives to construct efficient marketplaces to attract business, and in the interest of blockchain designers, who want their chains to be competitive. 
Facts to consider: A block-updating rate of multiple seconds will, by default, be slow compared to what legacy finance (LeFi) can offer, and the same is true for limits in on-chain transaction throughput. 
These two constraints are structural and hard to overcome. What may be fixable is the reordering of transactions by block builders, which can be guided towards a more beneficial outcome. 
MEV could, for example, be marginally impeded if the order of transactions in a block is fixed by a checkable deterministic algorithm. 
A deterministic ordering is clearly harder to manipulate than leaving the choice completely up to block builders. Unpredictability can be introduced either if the ordering is a function of components revealed as late as possible or given by a hard-to-invert function. 
It would be helpful, therefore, to have a one-way function that allows checking that the right process was carried out but makes predicting the block position fiendishly difficult.  To achieve this, imagine a one-way function that takes all the proposed transactions and splits out an `unstable ordering', i.e. if the transactions are slightly changed the resulting order changes dramatically.  To do this, calculate, for example, a hash function and use it to force an order.
Even this would offer only limited protection against sandwich or other attacks since if a particular combination of transactions does not give the right `sandwich' ordering, the block builder can try other similar block constructions. A particular order of three transactions, assuming that the rest of the block is neutral, is just one of six cases. 
If, instead, the aim is to combine dozens of transactions in a specific order, then the combinatorial possibilities multiply, and it becomes significantly harder for the block builder to deliver a particular outcome in the limited time available.

The main concern should instead not be the reordering or adding of transactions but rather the deliberate exclusion, which results in delays with costly economic consequences. As others have mentioned, this can lead to liquidation of positions as margin payments\footnote{ In the margining case, the trade-off between liquidation delays and risk for counterparties requires some new ideas. For example, can cross-chain links be utilized to allow margins to be replenished more flexibly? } are not received in time. This is especially worrisome at times of price volatility linked with natural chain congestion.      
Centralisation facilitates this process, 
since to motivate a block builder to disallow a transaction, a payment proportional to the associated `priority fee' is required. This cost will eventually outweigh any gain for a manipulator if it has to be paid out too many times. 

Therefore, collusion is the likely consequence  of restricting block building to a small group since, as Adam Smith wrote, ``people of the same trade seldom meet together, even for merriment and diversion, but the conversation ends in a conspiracy against the public."

\noindent
One of the authors - HCWP - thanks Numaan Ahmed for stimulating discussions, and
particular thanks to  Jim McDonald for clarifying some less-known aspects of the Ethereum blockchain. 

\clearpage

\begin{appendices}

\section{Estimation Attempt}
In this appendix, gas price data from Dune Analytics (\url{https://dune.com/}) was analyzed.
The logarithm of the median gas prices was taken, and a normal distribution was fitted. The Hurst exponent was estimated using the `hurst` package \footnote{\url{https://github.com/Mottl/hurst}}. The estimated Hurst exponent was $H \sim 0.38$.

\begin{table}[ht]
    \centering
    \caption{Summary Statistics}
    \label{tab:summary_stats}
    \begin{tabular}{lr}
        \toprule
        Statistic & Value \\
        \midrule
        Count & 1284938 \\
        Mean & 35.31 \\
        Standard Deviation (std) & 85.39 \\
        Minimum (min) & 2.00 \\
        25th Percentile (Q1) & 14.51 \\
        Median (Q2) & 23.15 \\
        75th Percentile (Q3) & 39.99 \\
        Maximum (max) & 18479.78 \\
        Skewness & 66.32 \\
        Kurtosis & 6643.85 \\
        \midrule
        Log-Normal Fit $\mu$ & 3.20 \\
        Log-Normal Fit $\sigma$ & 0.79 \\
        AIC & 11268105.19 \\
        BIC & 11268129.33 \\
        \bottomrule
    \end{tabular}
\end{table}

The Ornstein-Uhlenbeck (OU) process parameters were next estimated using linear regression. The estimated parameters, rounded to 3 significant figures, were 
$\kappa: 0.00700$, $\mu: 3.20$, and $\sigma: 0.0937$. From 2021-12-28 to 2024-06-10. The data was then used to simulate a fractional Ornstein-Uhlenbeck (FOU) process to enable memory effects with a Hurst exponent. The simulation of the FOU  relied on the `fbm` package \footnote{\url{https://github.com/crflynn/fbm}}. The Hurst exponent \(H\) was set to 0.5 for the simulation. Parameters \(\kappa\), \(\mu\), and \(\sigma\) were estimated through linear regression on the log-transformed data. The fractional Brownian motion (fBm) was generated using the Davies-Harte method, and the FOU process was simulated by solving the stochastic differential equation with the generated fBm. 

Using the estimated parameters, a FOU process was simulated. The simulation involved generating a fractional Brownian motion and using it to drive the OU process.
The histogram of the log of the median gas prices was plotted along with the fitted log-normal distribution and the FOU simulation. In  Figure 6, the reader can compare the observed data and the modelled distributions. The figure suggests that while the simplified FOU model employed captures some aspects of the data, it does not fully replicate the observed distribution.

\begin{figure}[h]
    \centering
    \includegraphics[width=0.8\textwidth]{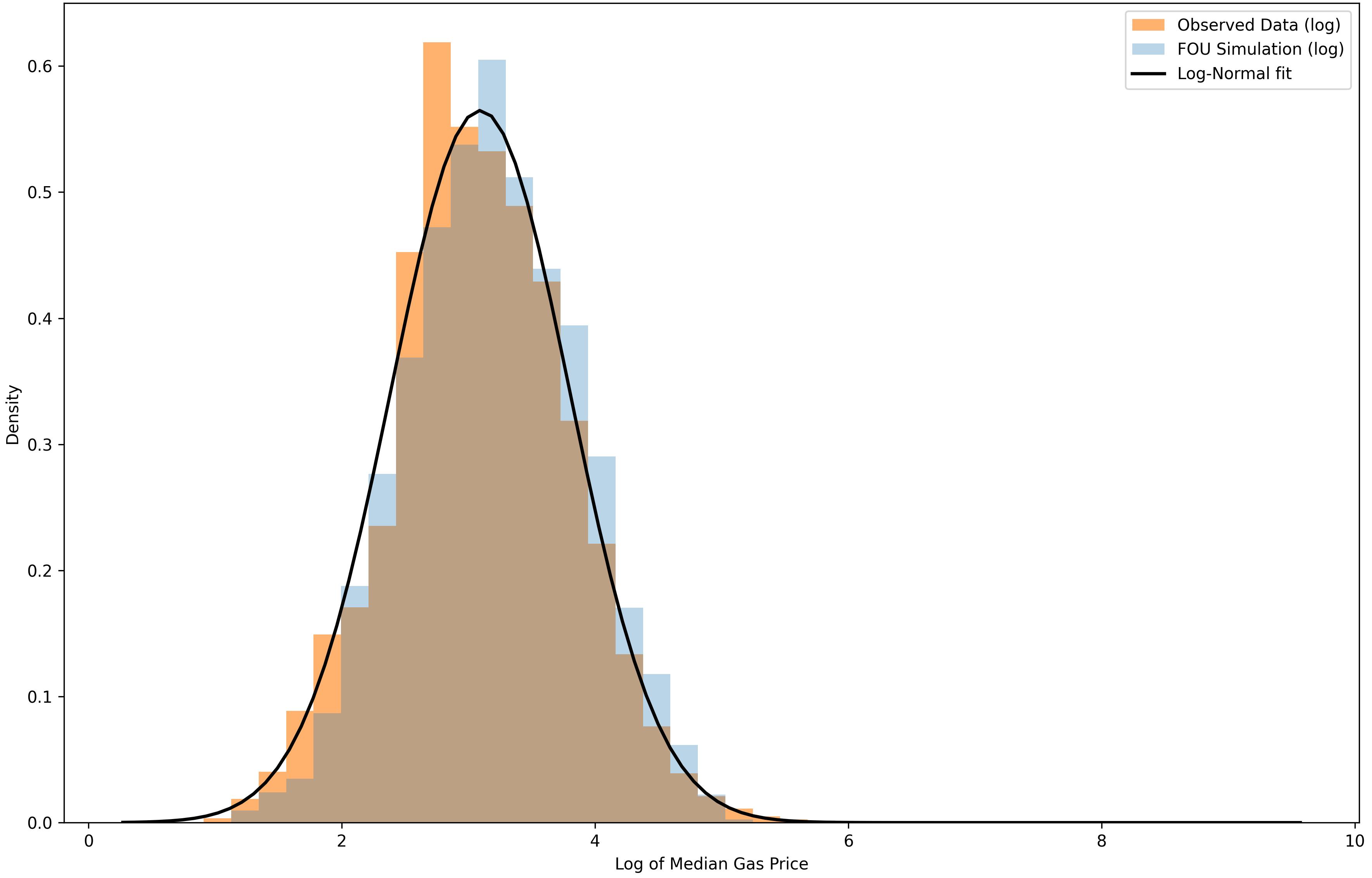}
    \caption{Histogram of Log Median Gas Prices with FOU and Log-Normal Fits. The plot shows the observed data in log space, the log-normal fit, and the FOU simulation. 
    The estimated parameters, rounded to 3 significant figures, were \(\kappa:  0.00700\), \(\mu: 3.20\), and \(\sigma:  0.0937\). H is set to $0.5$.
    }
    \label{fig:gas_price_fit}
\end{figure}

\clearpage

\section{Characteristics of Daily Gas Pricing}

In this section, we use again data from Dune Analytics (\url{https://dune.com/}) to study the trends in median gas prices. The data is in minute-level granularity, initially in Coordinated Universal Time (UTC) and converted to US Eastern Time (ET) to align with typical US market hours.  Year, date, and hour were extracted from each time stamp. Subsequently, the hourly median gas prices with error bars were determined to capture intra-day variability, as well as the daily median gas prices to observe general trends.
\begin{figure}[h!]
    \centering
    \includegraphics[width=\textwidth]{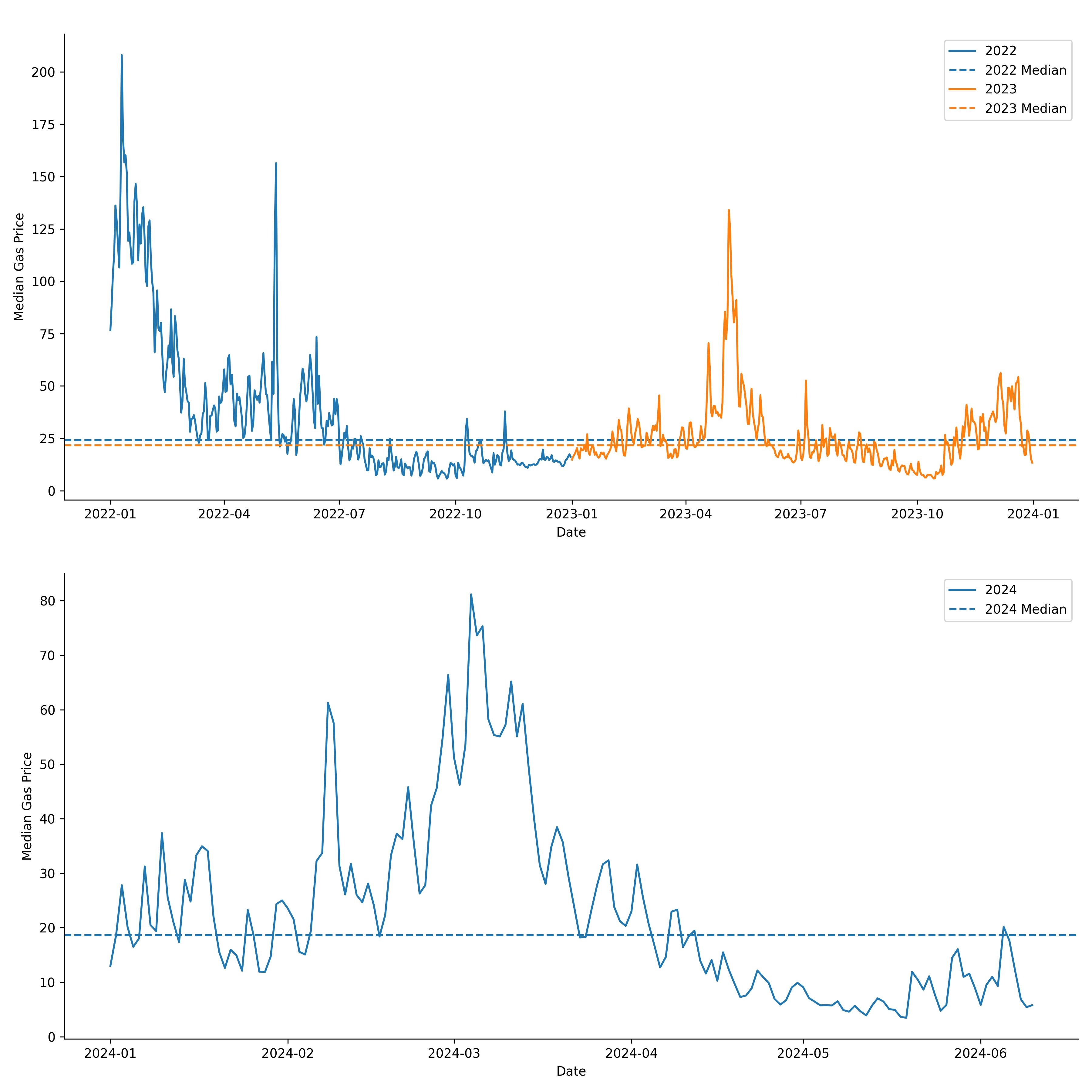}
    \caption{Daily Median Gas Prices are given by a solid line, years 2022 to 2023 (Top) and 2024 (Below).  The horizontal dashed lines represent the median gas prices for each year.}
    \label{fig:daily_median}
\end{figure}
The hourly median gas prices were calculated with error bars (standard deviation) indicating the standard deviation for each hour. The US market open (9:30 AM ET) and close (4:00 PM ET) times are marked with vertical dashed lines.

\begin{figure}[h!]
    \centering  \includegraphics[width=0.85\textwidth]{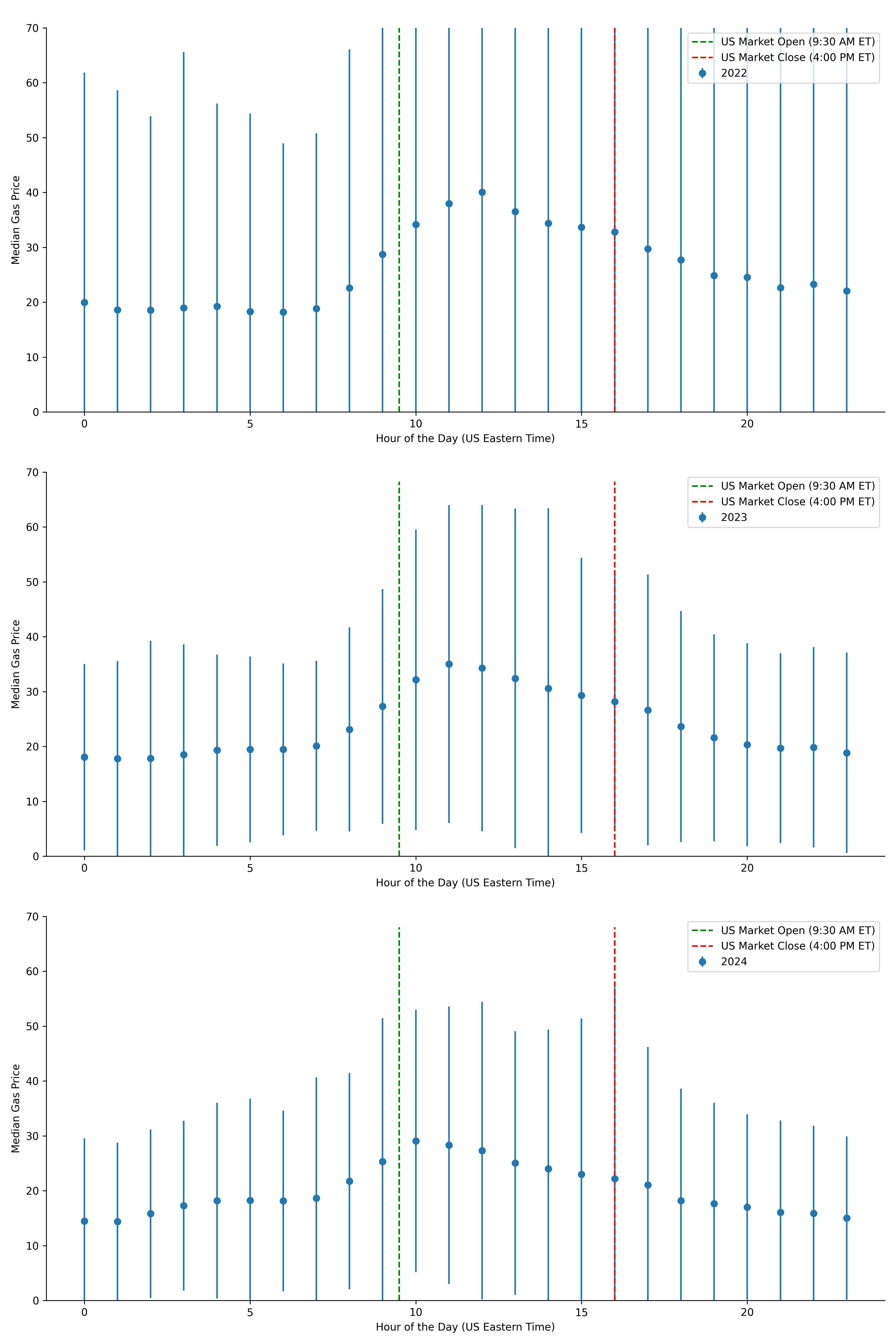}
    \caption{Hourly Median Gas Prices for 2022 (Top), 2023 (Middle), and 2024 (Bottom). The error bars are larger for 2022 and decrease over time, reflecting a pricing pattern that aligns with US trading hours. The vertical dashed lines indicate the US market open (green) and close (red) times.}
    \label{fig:hourly_median}
\end{figure}

The analysis shows the trends in median gas prices over different time frames. The hourly plots highlight the intra-day variations and the impact of US market hours, which appear to coincide with typical trading hours. while the daily plots provide an overview of the price trends over the years. The error bars reduce from 2022 to 2024, possibly showing the stabilizing effect of the implemented changes.

\clearpage

\section{Ethereum Improvement Proposals  and Their Effects on Gas Fees}

This section covers some of the specific influences the recent Ethereum Improvement Proposals 
 (EIPs) had on Gas Fees. 
In October 2020, EIP-2929 was introduced, increasing gas fees for certain state access opcodes to mitigate spam attacks and thereby to protect the network. It had mixed effects on overall cost. In August 2021, the London Upgrade, known as EIP-1559, introduced a base fee mechanism to make gas prices more predictable, helping to stabilize gas fees and as a consequence reduce  volatility. 
Later, in December 2021, EIP-4488 was released to reduce the cost of `calldata', decreasing gas costs for rollup solutions and effectively lowering fees for layer-2 solutions.

Most recently, on March 13 2024, the Dencun upgrade became active. 
It included several  EIPs to enhance
the network’s scalability and efficiency, focusing on layer-2 solutions.
A key component of the Dencun upgrade is proto-dank sharding (EIP-4844), which
introduced temporary data storage ``blobs'' to distribute some of the validating work.
This mechanism has significantly reduced transaction costs on layer-2 networks, which in
some instances led to gas fee reductions of over 90\%. As a result, median transaction
fees on layer-2 networks have reduced substantially. An increased migration of users and applications to these layer-2 networks
is anticipated,  reducing the load on the Ethereum mainnet and contributing
to lower overall gas prices.

\end{appendices}

\end{document}